\documentclass[preprint,aps,amsmath,superscriptaddress,nofootinbib,tightenlines]{revtex4}
\usepackage{bm}
\usepackage{epsfig}

\newcommand{\nn}{\nonumber} 

\newcommand{\bea}{\begin{eqnarray}}
\newcommand{\eea}{\end{eqnarray}}

\newcommand{\Pb}{\bar{P}}
\newcommand{\Vb}{\bar{V}}
\newcommand{\epsi}{\vec{\epsilon}^{\,\,*}_\psi}

\newcommand{\eD}{\vec{\epsilon}_{D^*}}

\newcommand{\egam}{\vec{\epsilon}^{\,\,*}_\gamma}
\newcommand{\vk}{\vec{k}}
\newcommand{\X}{X(3872)}

\begin{document}



\title{Radiative Decays $\boldmath X(3872) \to\psi(2S)\gamma$ and $\boldmath \psi(4040) \to X(3872) \gamma$
in Effective Field Theory}

\author{Thomas Mehen\footnote{Electronic address: mehen@phy.duke.edu}}
\affiliation{Department of Physics, 
	Duke University, Durham,  
	NC 27708\vspace{0.2cm}}

\author{Roxanne Springer\footnote{Electronic address: rps@phy.duke.edu}}
\affiliation{Department of Physics, 
	Duke University, Durham,  
	NC 27708\vspace{0.2cm}}

\date{\today\\ }


\begin{abstract}
Heavy hadron chiral perturbation theory (HH$\chi$PT) and XEFT are applied to the  decays $\X \to \psi(2S) \gamma$ and $\psi(4040)\to \X \gamma$ under the assumption that the $\X$ is a molecular bound state of neutral charm mesons. In these decays the emitted photon energies are 181 MeV and 165 MeV, respectively, so HH$\chi$PT can be used to calculate the underlying $D^0\bar{D}^{0*}+\bar{D}^0 D^{0*}  \to \psi(2S) \gamma$ or  $\psi(4040) \to (D^0\bar{D}^{0*}+\bar{D}^0 D^{0*} )\gamma$ transition. These amplitudes are matched onto XEFT to obtain decay rates. 
The decays receive contributions from both long distance and short
distance processes. We study the polarization of the $\psi(2S)$ in the decay  $\X \to \psi(2S) \gamma$ and the angular distribution of $\X$ in the decay  $\psi(4040)\to \X \gamma$ and find they can be used to differentiate between different decay mechanisms as well as discriminate between $2^{-+}$ and $1^{++}$ quantum number assignments of the $\X$.

\end{abstract}

\maketitle

\newpage

The $\X$~\cite{Choi:2003ue,Acosta:2003zx, Abazov:2004kp} is the first of many recently discovered hadrons containing
hidden charm that do not fit neatly into the traditional model of charmonia as nonrelativistic bound states of $c\bar{c}$. 
 The extreme closeness of the $\X$ to the $D^0 \bar{D}^{0*}$ threshold
has prompted many authors to suggest that the $\X$ is a molecular bound state of neutral charm mesons, though 
other possibilities including tetraquark interpretations have also been considered in the literature.  
For reviews of the recent discoveries in charmonium spectroscopy, 
see Refs.~\cite{Drenska:2010kg,Godfrey:2008nc,Voloshin:2006wf}. 

In this paper, we will work under the assumption that the $\X$ is a shallow $S$-wave bound state of $D^0\bar{D}^{0*} + \bar{D}^0 D^{0*}$ and calculate the 
radiative decays $\X \to \psi(2S) \gamma$ and $\psi(4040) \to \X \gamma$. The interpretation of the $\X$ as a charm meson molecule is motivated by the following considerations:
The observed branching ratios~\cite{Abe:2005ix}
\bea
\frac{\Gamma[\X\to J/\psi \pi^+ \pi^-\pi^0]}{\Gamma[\X \to J/\psi \pi^+ \pi^-]} =1.0 \pm 0.4\pm 0.3 \, ,
\eea
and \cite{delAmoSanchez:2010jr}
\bea
\frac{\Gamma[\X\to J/\psi \omega]}{\Gamma[\X \to J/\psi \pi^+ \pi^-]} =0.8\pm 0.3\, ,
\eea
indicate that the $\X$ couples with nearly equal strength to $I=0$ and $I=1$ final states. This rules out a conventional charmonium interpretation.
The observation of the decay $\X \to J/\psi\gamma$ demands $C = +1$ and the invariant mass distribution
in the decay $\X\to J/\psi \pi^+ \pi^-$ is consistent with the quantum number assignments $J^{PC} = 1^{++}$ or $2^{-+}$ only.
The decays $\X \to D^0\bar{D}^0\pi^0$ and $\X \to \psi(2S)\gamma$ would suffer an angular-momentum
suppression if the $J^{PC} = 2^{-+}$ assignment is correct, leading to a preference for $J^{PC}=1^{++}$.
If the quantum numbers of the $\X$ are $1^{++}$, then the $\X$ has an S-wave coupling to the $D^0\bar{D}^{0*} + \bar{D}^0 D^{0*}$. Finally, since the mass of the $\X$ is  $0.42 \pm 0.39$ MeV below the $D^0 \bar{D}^{0*}$ threshold~\cite{Brambilla:2010cs}, the $\X$ can mix strongly with $D^0\bar{D}^{0*}+\bar{D}^0 D^{0*}$ and the long range part of the $\X$ wavefunction should
be dominated by the $D^0\bar{D}^{0*}+\bar{D}^0 D^{0*}$ state.
Recently, the Babar collaboration studied the three-pion mass distribution in the  decay $\X \to J/\psi \pi^+ \pi^- \pi^0$ and concluded 
that the shape prefers the $2^{-+}$ assignment over $1^{++}$~\cite{delAmoSanchez:2010jr}.  However, the significance of their result is not so great that the $1^{++}$ 
assignment can be ruled out. The $2^{-+}$ assignment is problematic from the point of view of both the conventional
quark model as well other interpretations, for discussions see Refs.~\cite{Jia:2010jn,Burns:2010qq,Kalashnikova:2010hv}.
For the majority of this paper we will assume the $1^{++}$ assignment for the $X(3872)$ but we will also consider the implications of 
the $2^{-+}$ assignment for the radiative decays we will calculate below. An important point of this paper is that these observables may be able to
discriminate between the $1^{++}$ and $2^{-+}$ quantum number assignments.

If the $\X$ is indeed a shallow bound state of neutral charm mesons, then one can exploit the universal behavior of shallow bound states to compute many  $\X$ properties.
Universal quantities are those which depend only on the asymptotic form of the bound state wavefunction
and known properties of the constituents in the bound state. Examples of this for the $\X$  include the decay rates $\Gamma[\X \to D^0 \bar{D}^0 \gamma]$ and 
$\Gamma[\X \to D^0 \bar{D}^0 \pi^0]$, first calculated by Voloshin in Refs.~\cite{Voloshin:2003nt,Voloshin:2005rt}.
In the $X(3872)$, the wavefunction  of the $D^0\bar{D}^{0*}+\bar{D}^0 D^{0*}$ at a distance much greater than $R$, where $R$ is the range of the interaction between the charm mesons, takes on the form dictated by quantum mechanics, 
\bea\label{ert}
\psi_{DD^*}(r) \propto \frac{e^{-\gamma r}}{r} \, ,
\eea 
where $\gamma=\sqrt{2 \mu_{DD^*} B}$, $\mu_{DD^*}$ is the reduced mass of the $D^0$ and $\bar{D}^{0*}$, and $B$ is the binding energy. 
From the known binding energy, $B= 0.42\pm 0.39$ MeV, we infer a mean separation $r_X = 4.9^{+13.4}_{-1.4} \,$ fm, 
which is incredibly large compared to all known hadrons. 
 
Effective field theory offers a systematic approach to understanding the $\X$ as a molecule. The interactions of the theory are constrained by heavy quark and chiral symmetry via heavy hadron chiral perturbation theory (HH$\chi$PT)~\cite{Wise:1992hn,Burdman:1992gh,Yan:1992gz}.  XEFT~\cite{Fleming:2007rp} is a low energy effective
field theory of nonrelativistic $D^0, D^{0*}$, $\bar{D}^0$, $\bar{D}^{0*}$, and $\pi^0$ mesons near the $D^0 \bar{D}^{0*}$ threshold that is obtained from HH$\chi$PT by integrating
out virtual states whose energies are widely separated from the $D^0 \bar{D}^{0*}$ threshold.
At leading order (LO) XEFT reproduces the universal predictions that follow from the wavefunction in Eq.~(\ref{ert}). 

In Ref.~\cite{Canham:2009zq} elastic $D^{(*)}\X$ scattering was calculated using XEFT, and recently Ref.~\cite{Braaten:2010mg} applied 
XEFT to inelastic $\pi^+ \X$ scattering. Both these leading order calculations make predictions which depend only on the 
binding energy of the $\X$ and known properties of charm mesons with no other undetermined parameters. 
XEFT can also be used to systematically calculate corrections to universal predictions from effective range corrections,
other effects due to higher dimension operators in the XEFT Lagrangian,  and corrections from pion loops. 
In Ref.~\cite{Fleming:2007rp}, XEFT was used to calculate corrections to effective range theory predictions for the process $\X\to D^0 \bar{D}^0 \pi^0$. It was shown that corrections from pion loops were quite small, justifying a perturbative treatment of pions in XEFT. 

Finally, XEFT can be used to analyze properties that are not universal but depend on short distance aspects of the $\X$. Here, one seeks
factorization theorems for decay rates and cross sections which separate long distance from short distance scales in the $\X$. Factorization theorems
for $\X$ production and decay were first obtained in Refs.~\cite{Braaten:2005ai,Braaten:2005jj,Braaten:2006sy}. In XEFT these theorems are obtained by 
matching  HH$\chi$PT
amplitudes onto XEFT operators, then using these operators to calculate
decays and production cross sections in XEFT. 
An example is the  calculation of the
hadronic decays $\X\to \chi_J\pi^0$ and $\X\to \chi_J \pi \pi$~\cite{Fleming:2008yn}. These decays are interesting 
because the relative rates to final states with different $\chi_{cJ}$ can be predicted using heavy quark symmetry~\cite{Dubynskiy:2007tj}.
 
 In this paper, we apply XEFT to the radiative decays $\X \to \psi(2S) \gamma$ and $\psi(4040)\to \X \gamma$. 
 The BaBar collaboration quotes the branching fraction~\cite{:2008rn}:
\bea\label{br}
\frac{\Gamma[\X \to \psi(2S)\gamma]}{\Gamma[\X \to J/\psi \gamma]}  = 3.4 \pm 1.4 \, .
\eea
Later Belle searched for the decay $\X\to \psi(2S)\gamma$ but did not observe it and obtained an upper bound for the branching ratio
in Eq.~(\ref{br}) of $2.1$ with a confidence level of 90\%. This is consistent with Eq.~(\ref{br}) given the uncertainties, but suggests the true value 
may be lower than the central value in Eq.~(\ref{br}). 
Ref.~\cite{:2008rn} concluded that their measurement disfavored a molecular interpretation of the $\X$, largely because
 the branching ratio in Eq.~(\ref{br}) was predicted to be $3.7 \times10^{-3}$ in the specific molecular model of the $\X$ in Ref.~\cite{Swanson:2006st}. 
However, the ratio is sensitive to short-distance components of the $\X$ wavefunction which may not be modelled correctly in the model of Ref.~\cite{Swanson:2006st}.  
Ref.~\cite{Dong:2009uf} describes a model of the $\X$ as a mixed molecule-charmonium state that can account for   the branching ratio in Eq.~(\ref{br}).

XEFT alone will not yield a prediction for the branching fraction in Eq.~(\ref{br}) . Since the charm mesons must come to a point to coalesce into a quarkonium,  each absolute decay rate in the ratio is sensitive to short distance physics not described by XEFT.  Typically one would want to calculate ratios in which this short distance component cancels.  But  the $\psi(2S)$ and $J/\psi$ are members of different heavy quark multiplets, with couplings
unrelated by symmetry.  Finally, the photon energy in the decay $\X \to \psi(2S)\gamma$ is  181 MeV, which is within the range of applicability
of HH$\chi$PT, while the photon energy in the decay $\X \to J/\psi \gamma$ is 697 MeV, well outside the range of HH$\chi$PT.  So instead we will analyze what HH$\chi$PT and XEFT {\sl can} tell us about $\X \to \psi(2S) \gamma$.
We find that  there are two distinct mechanisms for the decay $\X \to \psi(2S) \gamma$ and that the polarization of $\psi(2S)$ will shed light on the relative
importance of these mechanisms. The polarization is calculated under both the  $J^{PC}$ =$1^{++}$ and $2^{-+}$ assumptions for the quantum numbers of the $\X$ and we discuss how this might be used to distinguish between them.

Another decay that can be analyzed in XEFT is $\psi(4040)\to \X \gamma$, in which the photon energy is 164 MeV. It may be possible to observe this decay at an $e^+e^-$ collider experiment such as BES III if the energy is tuned to the $\psi(4040)$ resonance. The angular distribution (relative to the beam axis) of the $\X$
produced in the process $e^+ e^- \to \psi(4040)\to \X \gamma$, yields similar information about the $\X$.

\section{$\X \to \psi(2S) \gamma$}

The procedure for calculating $\X$ decays to charmonium is described in detail in Ref.~\cite{Fleming:2008yn}. 
For the decay $\X \to \psi(2S) \gamma$, one first calculates the transition amplitude for $D^0 \bar{D}^{0*} + \bar{D}^0 D^{0*}\to \psi(2S)\gamma$
using HH$\chi$PT, extended to include charmonium states as explicit degrees of freedom. HH$\chi$PT Lagrangians 
with  quarkonia were first developed in Refs.~\cite{Casalbuoni:1992fd,Casalbuoni:1992yd}. For a recent application
to radiative decays of quarkonia, see Ref.~\cite{DeFazio:2008xq}. These papers used a covariant 
formulation in which the heavy mesons in the initial and final states can have distinct four-velocities. We will
use the two-component version of HH$\chi$PT introduced in Ref.~\cite{Hu:2005gf}. This formalism uses two-component spinors
with the four velocity for both the initial and final heavy mesons fixed to be $v^\mu=(1,\vec{0})$. This formalism is suitable for processes in
which the recoil of the heavy particle in the final state can be neglected, which is the case for this decay since  $v_i\cdot v_f = (m_X^2 +m_\psi^2)/(2 m_X m_\psi) =1.001$,
where $v_i (v_f)$ denotes the four-velocity of the initial (final) quarkonium.

\begin{figure}[t]
 \begin{center}
 \includegraphics[width=5.0in]{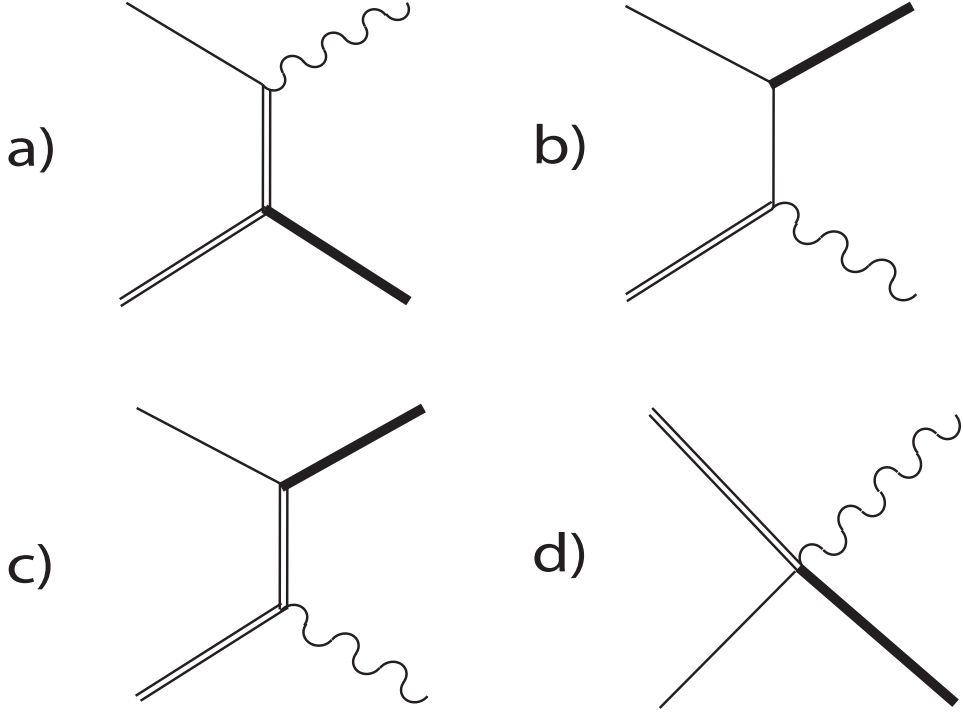}
 \end{center}
 \vskip -0.7cm \caption{Feynman diagrams contributing to the 
$D^0 \bar{D}^{0*} \to \psi(2S) \, \gamma$ 
amplitude. The thin solid line is a $D^0$ meson, the
double line is a $\bar{D}^{0*}$ meson, the wavy line is a photon, and the thick solid line
is the $\psi(2S)$.}
  \label{ddchipi}
 \end{figure}
 
The interaction Lagrangian for $\X \to \psi(2S) \gamma$ is given by
\bea\label{intL}
{\cal L} &=& \frac{e\beta}{2}{\rm Tr}[H_1^\dagger H_1 \, \vec{\sigma}\cdot \vec{B} \, Q_{11}] 
+ \frac{e Q^\prime}{2 m_c}{\rm Tr}[H_1^\dagger  \, \vec{\sigma}\cdot \vec{B} \, H_1] + h.c. \nn \\
&&+ i \frac{g_2}{2}{\rm Tr}[J^\dagger H_1 \vec{\sigma} \cdot  \stackrel{\leftrightarrow}{\partial} \bar{H}_1] 
+ \,i \frac{e c_1}{2}{\rm Tr}[J^\dagger H_1 \vec{\sigma} \cdot \vec{E} \bar{H}_1]  +h.c. \, .
\eea
Here  $Q_{11}=\frac{2}{3}$, $Q^\prime = \frac{2}{3}$, h.c. means hermitian conjugate, and $J$ is the superfield containing the $\psi(2S)$ and $\eta_c(2S)$.
The first two terms contain the couplings of the charm mesons to photons, the third term contains the coupling of the charm mesons 
to charmonia, and the final (contact) term  couples the charm mesons, the charmonia, and the electric field.
While $g_2$ and $c_1$ are unknown parameters at present,
$\beta$ occurs in HH$\chi$PT predictions involving measured quantities:
Ref.~\cite{Stewart:1998ke} obtains $\beta^{-1} \sim 1200$ MeV from radiative decays within the lowest charm meson multiplet,  Ref.~\cite{Hu:2005gf} found $\beta^{-1} = 275-375$ MeV,
and Ref.~\cite{Springer:1900zz} included the effects of the excited charm meson multiplet
to find $\beta^{-1}$ = 670 MeV. 
Since we have integrated the excited charm mesons out  and neglected loop corrections in the HH$\chi$PT calculations in this paper,
we will use the value of $\beta^{-1} = 275-375$ MeV extracted in Ref.~\cite{Hu:2005gf}, which makes the same approximations.
From these interactions we find four tree-level diagrams contributing to $D^0\bar{D}^{0*} +\bar{D}^0 D^{0*} \to \psi(2S) \gamma$,
which are shown in Figs.~\ref{ddchipi}a)-d). The amplitudes corresponding to each of these diagrams are 
\bea\label{amps}
a) &=& -\frac{g_2 \, e \,\beta_+}{3}\frac{1}{E_\gamma+\Delta}(  \vk \cdot  \epsi \,  \eD \cdot \vk \times \egam 
- \vk \cdot  \eD \,  \epsi \cdot \vk \times \egam) \\
b) &=& \frac{g_2 \, e \,\beta_+}{3}\frac{1}{\Delta -E_\gamma } \,  \vk \cdot  \epsi \,  \eD \cdot \vk \times \egam \\
c) &=& \frac{g_2 \, e \,\beta_-}{3}\frac{1}{E_\gamma } \,  \vk \cdot  \eD \,  \epsi \cdot \vk \times \egam \\
d) &=& -e \,c_1\, E_\gamma \, \eD \cdot \epsi \times \egam \, ,
\eea
where $\beta_\pm = \beta \pm 1/m_c$,  the polarization
vectors of the photon, $D^{0*}$, and $\psi(2S)$ are $\egam, \eD$, and
$\epsi$, respectively, and $\vec k$ is the outgoing photon momentum..  

 An additional potential contribution to $D^0 \bar{D}^{0*} +\bar{D}^0 D^{0*}
\to \psi(2S) \gamma$ is $D^0 \bar{D}^{0*}+ \bar{D}^0 D^{0*}\to \chi_{c1}(2P) \to \psi(2S) \gamma$.
 It is quite likely that the 
masses  of the $\chi_{cJ}(2P)$ states are close to the $\X$ mass. For example, Ref.~\cite{Barnes:2009zza} quotes quark model predictions for the $\chi_{c1}(2P)$ mass of 3925 MeV (in a nonrelativistic potential model) and 3953 MeV (in the Godfrey-Isgur relativistic quark model).
Alternatively, if the $Z(3930)$ is the $\chi_{c2}(2P)$ state one expects the $\chi_{c1}(2P)$
to be about 3885 MeV, assuming that the spin-orbit splitting for $\chi_{cJ}(2P)$ states is equal to the observed spin-orbit splitting for  $\chi_{cJ}(1P)$ states.  (The nonrelativistic  potential model predicts this splitting to 
be approximately the same, while the Godfrey-Isgur model predicts it to be slightly smaller.) In this scenario, the 
$\chi_{c1}(2P)$ is within 14 MeV of the $\X$ and the process $D^0 \bar{D}^{0*} \to \chi_{c1}(2P) \to \psi(2S) \gamma$
could be important for the radiative decay of the $\X$.

The decay $\chi_{cJ}\to \psi(2S) \gamma$ is an electric dipole transition mediated by the operator
\bea\label{vchi1}
{\cal L} = \delta^{2P2S} {\rm Tr}[J^\dagger \chi_c^i] E ^i  + h.c. \, ,
\eea
where $E^i$ is the electric field, $\chi_c$ is the super field containing the $\chi_{cJ}(2P)$ states, and the coupling constant 
$\delta^{2P2S}$ is the same as the one defined in Ref.~\cite{DeFazio:2008xq}, which calculated the  decay rate 
\bea
\Gamma[\chi_{c1}(2P)\to \psi(2S) \gamma] = \frac{(\delta^{2P2S})^2}{3 \pi}\frac{m_{\psi(2S)}}{m_{\chi_{c1}(2P)}} k_\gamma^3 \, .
\eea
 The charm mesons couple to the  
$\chi_{cJ}(2P)$ through a coupling
\bea\label{vchi2}
{\cal L} = \frac{i}{2} g_1^\prime \,{\rm Tr}[\chi_c^{\dagger i} \bar{H}\sigma^i H] +h.c. \, ,
\eea
This coupling is exactly the same as the coupling of heavy mesons to $\chi_{cJ}(1P)$ states introduced in Ref.~\cite{Fleming:2008yn}, except
now the $\chi^i_c$ superfield contains the $\chi_{cJ}(2P)$ states and the coupling is $g_1^\prime$ instead of $g_1$.
The effect of  including a tree-level diagram for $D^0\bar{D}^{0*} + \bar{D}^0 D^{0*}\to \psi(2S) \gamma$ using the vertices in Eqs.~(\ref{vchi1}) and (\ref{vchi2})
is to modify amplitude $d)$ in Eq.~(\ref{amps}) by the substitution
\bea 
e c_1 \to e c_1 + \frac{g_1^\prime \delta^{2P2S}}{m_X - m_{\chi_{c1}(2P)}} \, .
\eea
At present, $\delta^{2P2S}$, $g_1^\prime$, and $m_{\chi_{c1}(2P)}$ are unknown, so in what follows we will simply absorb this contribution into the definition
of the coupling $c_1$.

An illuminating observable is the decay rate for $\X \to \psi(2S)(\vec{\epsilon}_\psi)\gamma$, where  the polarization vector $\vec{\epsilon}_\psi$
of the produced $\psi(2S)$ can in principle be determined from the angular distribution of the leptons into which it decays: $\psi(2S) \to \ell^+ \ell^-$. 
Averaging over the initial $X(3872)$  and final photon polarizations we find
\bea \label{decayrate}
\Gamma[\X \to \psi(2S)(\vec{\epsilon}_\psi) \gamma] &=&\sum_\lambda |\langle 0| \frac{1}{\sqrt{2}}{\epsilon}^i(\lambda) 
\,(V^i \, \Pb +\Vb^i \, P) |X(3872,\lambda)\rangle|^2 \\
&& \times \frac{m_\psi}{m_X}\frac{E_\gamma}{24\pi} \left(  \frac{2}{3}\left( A +  C\right)^2\,|\hat{k} \cdot \vec \epsilon_\psi|^2 
+\frac{1}{3}\left( B-C\right)^2|\hat{k} \times \vec\epsilon_\psi|^2  \right) \nn \, ,
\eea
where $V^i$ and $P$ are the vector and scalar components of the $D^{(*)}$ superfield, and  $\epsilon^i(\lambda)$ are a basis
of polarization vectors for the $X(3872)$.  
%
%
%
In Eq.~(\ref{decayrate}), $\hat k$ is a unit vector in the direction of the photon's three-momentum, and
\bea
A = \frac{g_2 e \beta_+}{3} \frac{2 E^3_\gamma}{\Delta^2-E_\gamma^2} \quad
B=  \frac{g_2 e }{3} \frac{\beta_+ E^2_\gamma+ \beta_- E_\gamma (E_\gamma+\Delta)}{E_\gamma+\Delta} \quad
C= - e c_1 E_\gamma \, .
\eea
We have used 
$ \vec\epsilon_\psi^{\,*}\cdot \vec\epsilon_\psi =|\hat{k} \cdot \vec \epsilon_\psi|^2 
+ |\hat{k} \times \vec\epsilon_\psi|^2$.
The total decay rate is given by
\bea\label{tr}
\Gamma[\X \to \psi(2S) \gamma] &=& 
\sum_\lambda |\langle 0| \frac{1}{\sqrt{2}}{\epsilon}_i(\lambda) 
\,(V^i \, \Pb +\Vb^i \, P) |X(3872,\lambda)\rangle|^2\nn \\
&&\times \frac{E_\gamma}{36\pi}  \frac{m_\psi}{m_X} \left[(A+C)^2+(B-C)^2\right] \,.
\eea
In addition to not having an experimental determination of the
parameters $g_2$ and $c_1$ contained in $A$, $B$, and $C$, the
matrix element in Eq.~(\ref{tr}) is unknown; additional measurements
will be necessary to make a prediction for the total rate. However, the matrix element between $\X$ and its constituents appears
in any process involving the $\X$, so a measurement from a different
production or decay chain can be used in this calculation.
Combining the lower bound  $\Gamma[\X \to \psi(2S) \gamma] /\Gamma[\X]> 3.0 \times 10^{-2}$
from Refs.~\cite{:2008rn,pdg} with the upper bound on the total width $\Gamma[\X]<2.3$ MeV \cite{Choi:2003ue}
yields the lower bound on the  partial width $\Gamma[\X \to \psi(2S) \gamma] >7 \times 10^{-2}$ MeV.

If we define $|{\cal M}_\parallel |^2$ ($|{\cal M}_\perp |^2$ ) to be the matrix element squared for decay into 
$\psi(2S)$ polarized parallel (perpendicular)  to the axis defined by the photon momentum, then
\bea
|{\cal M_\parallel}|^2 &=&  \frac{2}{3}\left( A +  C\right)^2\nn \\
|{\cal M_\perp}|^2 &=&  \frac{2}{3}  \left( B-C\right)^2 \, .
\eea
It is interesting to consider the limits i) $|g_2 \beta_\pm | \ll |c_1|$ and ii) $|g_2 \beta_\pm| \gg |c_1|$.  When $|g_2 \beta_\pm |\ll |c_1|$
the short distance contribution dominates, $|C| \gg |A|,|B| $, and 
\bea
i) \qquad \frac{|{\cal M}_\parallel |^2}{|{\cal M}|^2}= \frac{|{\cal M}_\parallel |^2}{|{\cal M}_\parallel |^2+|{\cal M}_\perp |^2}= \frac{1}{2} \, .
\eea
That is, diagram d) yields  $|{\cal M}_\parallel|^2 = |{\cal M}_\perp|^2$. In case ii), diagrams a) -c) dominate and we find 
\bea\label{ii}
ii) \qquad  \frac{|{\cal M}_\parallel |^2}{|{\cal M}|^2}
= \frac{4 E_\gamma^4}{4 E_\gamma^4+(E_\gamma+ r_{\beta}(E_\gamma+\Delta))^2(E_\gamma-\Delta)^2} = 0.95 \,(0.92)\, .
\eea
where $r_\beta\equiv \beta_-/\beta_+$.
The first number on the r.h.s.~of Eq.~(\ref{ii}) corresponds to $r_{\beta}$ in the range  0.62-0.69, taken from fits in Ref.~\cite{Hu:2005gf},
while the number in parentheses corresponds to $r_{\beta} = 1$.
In case ii) diagrams a)-c) dominate over diagram d), and diagram
b) dominates diagrams a)-c) because $E_\gamma -\Delta \sim 39$ MeV is small.
The result is that the polarization of the produced  $\psi(2S)$ is dictated by diagram b), which peaks for longitudinally polarized $\psi(2S)$.
The angular distribution of the final state lepton pair in the decay $\psi(2S) \to \ell^+ \ell^-$ 
is 
\bea
\frac{d\Gamma}{d \cos\theta} \propto 1+\alpha \cos^2\theta \qquad \alpha =\frac{1-3 f_L}{1+f_L} \, ,
\eea
where $f_L = |{\cal M}_\parallel |^2/|{\cal M}|^2$ and $\cos\theta$ is the angle between the lepton's and the photon's momentum. 
For case i) $\alpha=-1/3$ and for case ii) $\alpha= -0.91 (-0.95)$, so the angular distribution of the leptons is sensitive to the production mechanism
and can be used to distinguish among them. 

Defining $\lambda =3c_1/(g_2 \beta_+)$,  $\lambda \to 0$ corresponds to
diagrams a)-c) dominating, while
$|\lambda| \to \infty$ corresponds to the contact interaction dominating. In terms of $\lambda$,
\bea
f_ L = \frac{N}{D} \, , \nn
\eea
where
\bea
N &=&\left(\frac{2 E^2_\gamma}{\Delta^2-E_\gamma^2}\right)^2 -\lambda \, \frac{4 E_\gamma^2}{\Delta^2-E_\gamma^2} + \lambda^2\\ 
D &=& \left(\frac{2 E^2_\gamma}{\Delta^2-E_\gamma^2}\right)^2+ \left(\frac{E_\gamma+r_\beta(E_\gamma+\Delta)}{E_\gamma + \Delta}\right)^2
-2 \lambda \, \left( \frac{2 E_\gamma^2}{\Delta^2-E_\gamma^2} -\frac{E_\gamma+r_\beta(E_\gamma+\Delta)}{E_\gamma+\Delta} \right) + 2 \lambda^2 \, .  \nn 
\eea
\begin{figure}[t]
 \begin{center}
 \includegraphics[width=4.0in]{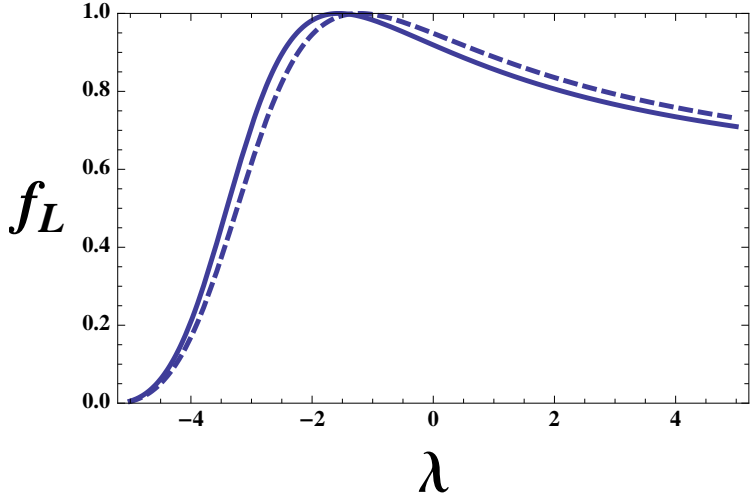}
 \end{center}
 \vskip -0.7cm \caption{$f_L$ as a function of the parameter $\lambda$ (defined in text). Solid line corresponds to $r_\beta=1.0$, dashed line 
 to $r_\beta=0.66$. } \label{fL}
 \end{figure}

\begin{figure}[t]
 \begin{center}
 \includegraphics[width=4.0in]{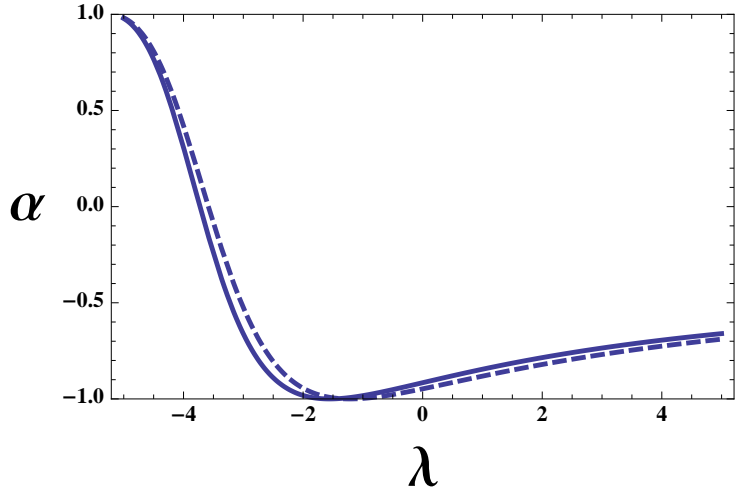}
 \end{center}
 \vskip -0.7cm \caption{$\alpha$ as a function of the parameter $\lambda$ (defined in text). Solid line corresponds to $r_\beta=1.0$, dashed line 
 to $r_\beta=0.66$.} \label{alpha}
 \end{figure}
Fig.~\ref{fL} is a plot of $f_L$ as a function of $\lambda$ and Fig.~\ref{alpha} is a plot of $\alpha$ in terms of the parameter $\lambda$. Naive dimensional analysis suggests $\lambda \sim O(1)$, so the plots
range over $-5< \lambda <5$.  The plots show the results for $r_\beta=1$ (solid) and $r_\beta=0.66$ (dotted). 
The behavior shown in the plots remains the same when $r_\beta$ is varied between
$ -1 < r_\beta < 1$, where the lower limit corresponds to the
situation where the $1/m_c$ term (cf. Eq.~(\ref{intL})) dominates while the upper limit
is the heavy quark limit.  The curves just continue to move to the right for smaller values of $r_\beta$.
For the most likely values of $r_\beta$, longitudinal polarization ($f_L\geq 1/2$ and $\alpha \leq -1/3$) is found for
$\lambda $ in the range $-3 < \lambda <5$.

This analysis potentially yields a method for determining
the amount of a molecular versus nonmolecular description consistent with
a $1^{++}$ assignment for the $\X$.  If the multipole expansion is legitimate, the leading order description of a nonmolecular $1^{++}$
is a $P$-wave contact term equivalent to $c_1$.  So to the extent that
the $\psi(2S)$ polarization in the $X(3872) \rightarrow \psi(2S)
\gamma$ decay is found to be longitudinally polarized, the molecular
description dominates its character.
 
It is also interesting to consider what the $J^{PC}=2^{-+}$ assignment for the $\X$ would imply 
for the $\psi(2S)$ polarization. Denote the spin-2 field in HH$\chi$PT by $X^{ij}$,  where $X^{ij}$ is symmetric and traceless 
in its indices. The simplest coupling mediating  $\X\to \psi(2S)\gamma$ is 
\bea \label{X2}
{\cal L} = g^\prime \,{\rm Tr}[X^{ij}J^\dagger\sigma^i] B^j \,,
\eea
which yields an amplitude proportional to 
\bea
{\cal M}[\X \to \psi(2S)(\vec{\epsilon}_\psi) \gamma] \propto \vec{\epsilon}_\psi^{\,*i} (\vec{k}\times \vec{\epsilon}_\gamma^{\,*})^j h^{ij}\, ,
\eea 
where $\vec{k}$ is the photon three-momentum, and $\vec{\epsilon}_\psi^{\,*} $, $\vec{\epsilon}_\gamma^{\,*}$, and $h^{ij}$ are the polarization tensors for the $\psi(2S)$, photon, and $\X$, respectively. 
Summing over the polarizations of the $\X$ and the photon,  the cross section's dependence on the polarization of the $\psi(2S)$ becomes
\bea
\sum \left|{\cal M}[\X\to \psi(\vec{\epsilon}_\psi) \gamma ] \right|^2 \propto |\vec{k} \cdot \vec{\epsilon}_\psi|^2 + \frac{7}{6} |\vec{k} \times \vec{\epsilon}_\psi|^2 \,.
\eea
The fraction of longitudinally polarized $\psi(2S)$ is $f_L=0.3$, corresponding to $\alpha = 0.08$. This leading order description
of a  $J^{PC} = 2^{-+}$ $\X$ yields a very slight transverse polarization of the $\psi(2S)$.

 Ref.~\cite{Jia:2010jn} assumes that the $\X$ is the $\eta_c({}^1D_2)$. In the models considered in that paper, the leading contribution
to the decay is an M1 amplitude identical in form to that given by Eq.~(\ref{X2}).  In addition, the models  include  electric quadrupole and magnetic octopole transitions (which correspond to higher dimension operators in
HH$\chi$PT). From the helicity amplitudes calculated in the five potential models of Ref.~\cite{Jia:2010jn}, we obtain $f_L=0.11-0.28$ ($\alpha= 0.13-0.6$). This suggests the $J^{PC}=2^{-+}$ quantum number assignment prefers slightly transverse polarization for the $\psi(2S)$ in the $\X \to \psi(2S) \gamma$ decay. In contrast, the molecular $(1^{++})$ hypothesis predicts longitudinal polarization in much (but not all) of  parameter space.

\section{$\psi(4040) \to \X \gamma$}

\begin{figure}[t]
 \begin{center}
 \includegraphics[width=4.0in]{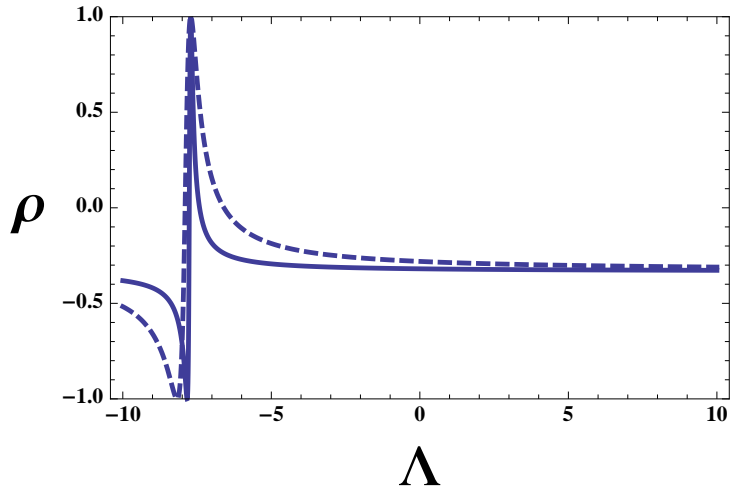}
 \end{center}
 \vskip -0.7cm \caption{The parameter $\rho$ from the angular distribution of $\X$ in the decay $\psi(4040)\to \X \gamma$
as a function of $\Lambda \equiv 3 \tilde c_1/(\tilde g_2 \beta_+)$. The solid line has $r_\beta$ = 1.0 and the dashed line has $r_\beta$ = 1.0.}
 \label{rho}
 \end{figure}

Assuming that the $\psi(4040)$ is the $3{}^3S_1$ charmonium,  the interaction Lagrangian  for $\psi(4040) \to \X  \gamma$ is essentially 
the same as in Eq.~(\ref{intL}). The superfield $J$ should be replaced by the superfield containing the $\psi(4040)$ while the couplings $g_2$ and
$c_1$ are replaced by analogous couplings $\tilde g_2$ and $\tilde c_1$. The diagrams for  $\psi(4040) \to (D^0 \bar{D}^{0*}+ \bar{D}^0D^{0*})\gamma$
are related to those in Fig.~\ref{ddchipi} by crossing symmetry. The corresponding amplitudes are 
\bea\label{4040amps}
a) &=& -\frac{\tilde g_2 \, e \,\beta_+}{3}\frac{1}{E_\gamma-\Delta}(  \vk \cdot  \vec{\epsilon}_\psi \,  \eD \cdot \vk \times \egam 
- \vk \cdot  \eD \,  \vec{\epsilon}_\psi\cdot \vk \times \egam) \\
b) &=&- \frac{g_2 \, e \,\beta_+}{3}\frac{1}{ E_\gamma +\Delta} \,  \vk \cdot  \vec{\epsilon}_\psi \,  \eD \cdot \vk \times \egam \\
c) &=& \frac{\tilde g_2 \, e \,\beta_-}{3}\frac{1}{E_\gamma } \,  \vk \cdot  \eD \,   \vec{\epsilon}_\psi \cdot \vk \times \egam \\
d) &=& -e \,\tilde c_1\, E_\gamma \, \eD \cdot \vec{\epsilon}_\psi \times \egam \, .
\eea
We are interested in the angular distribution of the $\X$ produced in the process  $e^+e^- \to \psi(4040) \to X(3872)\gamma$.
The mass of the electrons is negligible compared to the $\psi(4040)$; the electrons are treated as helicity eigenstates
whose spin angular momentum is projected along the beam axis.
Thus the $\psi(4040)$ has $L_z =\pm 1$, where the beam
axis defines the $z$-direction. Therefore the $\psi(4040)$ is produced with polarization normal to the beam axis. This then 
dictates the angular distribution of the $\X$ produced in the decay. If we square the amplitudes, and average over the $\X$ and $\gamma$
polarizations, we find the matrix element squared is 
\bea\label{msqr}
\sum |{\cal M}(\vec{\epsilon}_\psi)|^2 \propto \frac{2}{3} P\, |\hat{k}\cdot \vec{\epsilon}_\psi|^2 + \frac{1}{3} T\,  |\hat{k}\times \vec{\epsilon}_\psi|^2 \, ,
\eea 
where $\vec{\epsilon}_\psi$ is the $\psi(4040)$ polarization vector, $\hat{k}$ is unit-vector along the three-momentum of the photon in the $\psi(4040)$ rest frame, and 
$P$ and $T$ are given by:
\bea
P&=& \left( \frac{\tilde g_2 e \beta_+}{3}\frac{2 E_\gamma^3}{\Delta^2-E_\gamma^2} -e \tilde c_1 E_\gamma \right)^2 \nn \\
T &=&  \left(  \frac{\tilde g_2 e \beta_+}{3}\frac{E_\gamma^2+r_\beta E_\gamma (E_\gamma-\Delta)}{E_\gamma-\Delta} +e \tilde c_1 E_\gamma  \right)^2 \,.
\eea
The angular distribution can be obtained by replacing $\epsilon_\psi^i \epsilon_\psi^{*j} = \delta^{ij} -\hat{z}^i \hat{z}^j$ in Eq.~(\ref{msqr}). Defining $\theta$ to be the angle that the $\X$ (or the photon) makes with the beam axis, we find
\bea\label{ang}
\frac{d\sigma}{d \,\cos \theta} \propto 1+ \rho \cos^2\theta \, ,
\eea
where  $\rho$ is given by 
\bea
\rho = \frac{ T - 2P }{T  + 2P  } \, .
\eea
The value of $\rho$ in Eq.~(\ref{ang}) depends on  the following combination of HH$\chi$PT coupling constants:
\bea
\Lambda \equiv \frac{3 \tilde c_1}{\tilde g_2 \beta_+} \, .
\eea
Fig.~\ref{rho} is a plot of $\rho$ as a function of the dimensionless
parameter $\Lambda$, for $-10\leq \Lambda \leq 10$. $\Lambda$ is expected to be $O(1)$. In the region where $\tilde c_1$ dominates, $|\Lambda|  \rightarrow \infty$, and $\rho$ asymptotes to $-1/3$.   As $r_\beta$ decreases, $\rho$ reaches the asymptote at larger values of $\Lambda$.  Near $\Lambda \sim -8$, $\rho$ is very sensitive to $\Lambda$ and can take on any value between $-1$ and +1. For comparison, if the $\X$ has quantum numbers $J^{PC}=2^{-+}$ and couples to the $\psi(4040)$ and the photon by
the leading order operator analogous to Eq.~(\ref{X2}), $\rho = 1/13 = 0.08$.
So the angular distribution of $\X$ produced in the process $e^+ e^- \to \psi(4040) \to \X \gamma$
can also be used to discriminate between quantum number assignments of the $\X$.

\section{Summary}

In this paper we have calculated the radiative decays $\X \to \psi(2S)\gamma$ and $ \psi(4040) \to \X \gamma$ using  XEFT. 
Each receives contributions from a ``long-distance" portion involving
the propagation of a heavy charm meson (diagrams a)-c) in Fig.~(\ref{ddchipi})), and a short-distance contact operator (diagram d)
in Fig.~(\ref{ddchipi})).
The relative importance of these two types of  diagrams depends on the ratio of two undetermined parameters in the HH$\chi$PT Lagrangian; $\lambda$ for the $\X$ decay mechanism above and $\Lambda$ for the $\X$ production mechanism.
A primary result of this paper is that the angular distributions of
decay products can be used to distinguish between the $1^{++}$ and $2^{-+}$ assignments of the $X(3872)$ as well as the relative importance
of the two types of diagrams involved.  The polarization of the $\psi(2S)$ produced in the decay $\X_{1^{++}}\to \psi(2S)\gamma$ is sensitive to $\lambda$. In  much of the parameter space
the $\psi(2S)$ is longitudinally polarized. In contrast, for $\X_{2^{-+}} \to
\psi(2S) \gamma$, $\psi(2S)$ is produced with a slight transverse polarization.
A similar set of diagrams to those in Fig.~(\ref{ddchipi})) (with different coupling constants) contributes to the decay $\psi(4040) \to \X \gamma$.
In the process $e^+e^- \to \psi(4040) \to\X \gamma$, the angular distribution of the $\X$ (or $\gamma$) relative to the $e^+e^-$ beam axis  can discriminate
between the $1^{++}$ and $2^{-+}$ assignments of $\X$.
In most of parameter space, the parameter $\rho$ in Eq.~(\ref{ang}) is near $-1/3$ for $X(3872)_{1^{++}}$, while $X(3872)_{2^{-+}}$ produces $\rho \approx 0.08$.

\acknowledgments 

This work was supported in part by the  Director, Office of Science, Office of Nuclear Physics, of the U.S. Department of Energy under grant numbers 
DE-FG02-05ER41368, and DE-FG02-05ER41376.


\end{document}